\documentclass[11pt,english]{article}
\usepackage{hyperref} 
\pdfoutput=1
\begin{document} 
\vspace*{-1cm}
\begin{center}
	{\bf {\Large The trapping and escape of buoyant vortex rings in sharply stratified fluids}}\\
\vspace*{.5cm}
R. Camassa, S. Khatri, R. McLaughlin, K. Mertens, E. Monbureau, D. Nenon,\\  C. Smith, C. Viotti, B.White, \\Joint Fluids Laboratory, University of North Carolina at Chapel Hill
\end{center}
\begin{abstract}
In this fluid dynamics video we study the dynamics of {\it miscible} vortex rings falling in ambient strongly (near two-layer) stratified fluid. Experiments and direct numerical simulations using the variable density Navier-Stokes (VARDEN) solver are presented and compared.  Critical phenomena are identified depending upon the key parameters of the experiment (fluid and ring densities, upper layer vortex travel distance, etc) in which the descending dense vortex ring may experience complete trapping, partial trapping, or fissioning into a cascade of smaller vortices.  The interaction of the vortex ring with the upper layer fluids leads to viscous entrainment which alters the effective buoyancy of the ring.  Upon impinging on a density transition, the entrained fluid imparts different dynamics as it attempts to regain equilibrium leading to the critical behaviors.  
\end{abstract}


 In this study results are presented concerning the various phenomenological behaviors which can be exhibited when a dense vortex ring descends through a sharp density transition.
It may initially be surprising that a fluid of greater density than either of the two layers in an underlying tank can become trapped at the density transition. During the descent however a growing entrainment bubble of light upper layer fluid forms and travels downward with the vortex ring. Upon entering the lower layer fluid the surrounding entrainment bubble feels a buoyant force which will cause a rebound at some depth within the transition or lower layer. During this rebound process enhanced mixing occurs between the entrained upper layer fluid and the ring content which can allow for the conglomerate mixture to equilibrate to the local ambient density, causing the ring content to become trapped within the transition layer. In this video we consider a fixed top layer density of fresh water (0.998 g/cc) stably stratified above a fixed density of salt water (1.020 g/cc) with transition layer on the order of 1 cm. By varying the length scale of the top layer fluid (between 10-70 mm) and droplet density (between 1.020-1.040 g/cc) four phenomenological behaviors are observed. At large top layer thicknesses and small density differences the ring contents will become completely trapped within, or on top of, the transition layer (with possible damped oscillations), coined "settling". By lowering the top layer length scale and/or increasing the droplet density the behavior transitions into rings which can be destroyed upon impact with the transition layer. In this process fingers form and can potentially break through the transition layer, though the majority of ring content will still remain trapped within this transition layer, coined "chandeliers". If one further decreases the top layer length scale and/or further increases droplet density this behavior again transitions. In this situation the vortex ring and entrainment bubble pass below the transition layer, however when the entrainment bubble rebounds it carries the entire ring contents back to the transition layer where it remains due to the strong mixing during rebound, coined "bouncing". Lastly, if we continue to reduce the top layer thickness further and/or increase droplet density we eventually reach a situation in which the entrainment bubble can no longer carry the entire ring content back to the transition layer. While some of the ring content will rebound, material from the core of the vortex ring continues downward, coined "core-fallout".  Direct comparisons between the experiment and the DNS are presented for the "chandelier" case are presented.

\end{document}